\documentclass[structabstract]{aa}
\usepackage{graphicx}
\usepackage{epsfig}
\usepackage{color}
\usepackage{txfonts}
\usepackage[authoryear]{natbib}
\bibpunct{(}{)}{;}{a}{}{,}

\begin{document}

\title{A search for dormant binaries with degenerate components
in $\omega$~Centauri and NGC~6397}

\titlerunning{A search for degenerate binaries in $\omega$~Centauri and NGC~6397}

\subtitle{}

\author{M. Rozyczka\inst{1}
        \and
        J. Kaluzny\inst{1}
        \and
        P. Pietrukowicz\inst{1,2}
        \and
        W. Pych\inst{1}
        \and
        M. Catelan\inst{2}
        \and
        C. Contreras\inst{2}
        \and
        I.~B. Thompson\inst{3}
        }

\authorrunning{Rozyczka et al.}

\offprints{M. Rozyczka,\\ \email{mnr@camk.edu.pl}}

\institute{Nicolaus Copernicus Astronomical Center,
ul. Bartycka 18, 00-716 Warszawa, Poland
        \and
Pontificia Universidad Cat\'olica de Chile,
Departamento de Astronom\'ia y Astrof\'isica,
Av. Vicu\~na MacKenna 4860,
782-0436 Macul, Santiago, Chile
        \and
Carnegie Institution of Washington,
813 Santa Barbara Street, Pasadena, CA 91101, USA
}

\date{Received ...; accepted ...}

\abstract
{}
{
We report on the first spectroscopic search for quiescent degenerate
binaries in globular clusters.
}
{
Our survey is based on a sample of short-period optical variables
which are likely optical counterparts of quiescent X-ray sources
in $\omega$~Centauri (NGC 5139) and NGC 6397.
}
{
The studied candidates have nearly sinusoidal light curves with
amplitudes of 0.05--0.12~mag in $V$ (0.35 mag in one case), and periods
of 0.1--1.3~days. This type of variability, most probably originating
from the ellipsoidal effect, has been observed in X-ray novae when they
settled into quiescence after an outburst.
}
{
We find that two of the surveyed systems harbour dim components
with masses in excess of 1~$M_\odot$, making them attractive
targets for future investigations. We also suggest that there
are two subpopulations of blue stragglers in $\omega$~Centauri,
differing in mass-transfer history and/or helium content.
}

\keywords{globular clusters: individual: NGC~5139 ($\omega$~Centauri),
NGC~6397 -- binaries: close -- binaries: spectroscopic -- blue stragglers}

\maketitle

\section {Introduction}

Since the early surveys with the Uhuru and OSO-7 satellites it has been
known that X-ray sources occur about a thousand times more frequently in
globular clusters (GCs) than in the rest of the Galaxy \citep[e.g.,][]{jk75,gc75}.
Over fifteen hundred of them have been detected to date in these stellar systems
by recent missions such as Chandra and XMM-Newton
\citep[][ and references therein]{poo10}. In addition, over a hundred millisecond
pulsars~-- objects which have evolved in low-mass X-ray binaries~-- are
known to reside in GCs \citep{lyn10}. Most of the X-ray sources are
binaries with degenerate components (neutron stars or white dwarfs),
henceforth referred to as ``degenerate binaries''. Remarkably, no
stellar-mass black holes (BH) are known in Galactic GCs, despite
theoretical predictions that they should exist \citep[e.g., ][]{dev07}.
The presence of intermediate-mass black holes (IMBHs)
($100\,M_{\odot}< M <10^{3-4}~M_{\odot}$) in GCs has also not yet been proven
\citep[e.g.,][]{mac08,van10}.

While the origin of IMBHs is unclear, stellar-mass black holes
should form along with neutron stars via supernova explosions during
the early evolution of a cluster. The very presence of such objects
would {\bf provide} interesting information concerning the
effectiveness of dynamical interactions leading to the ejection of
BHs into the intracluster medium, which according to the predictions
of current approximate models should be very high
\citep[e.g.,][]{dow10}. Their absence would confirm those
predictions. Alternatively, it could mean that massive
low-metallicity stars cannot produce BHs, which would have important
implications for the origin of early BHs, believed to be seeds for
the first galaxies in the Universe. In either case, searching for
black holes in GCs is certainly a worthwhile task.

It is well known that field X-ray novae (binaries most probably
hosting stellar-mass black holes) spend most of their time in
quiescence, showing only ellipsoidal variability in the visible
domain \citep{rem06}. Since GCs harbor a multitude
of degenerate binaries, they must also contain many systems
which presently accrete at a very low rate or do not accrete at all,
thus being very weak or undetectable in the X-rays. Based on X-ray
luminosity ($L_{\mathrm X}$) and hardness ratio, Grindlay (2006)
splits the population of weak X-ray sources in GCs into four
major classes. The first three, arranged according to $L_X$
increasing from $10^{29}$ (the present sensitivity  limit for
the nearest GCs) to $10^{32}$ erg s$^{-1}$ are i) active binaries
(binary main-sequence stars in which the X-ray emission
originates from chromospheric activity, e.g. BY~Dra type systems);
ii) cataclysmic variables (in which a white dwarf accretes from a
low-mass main-sequence companion); iii) quiescent low-mass X-ray
binaries (in which a neutron star intermittently accretes from a
main-sequence or evolved companion), hereafter referred to as
qLMXBs. The fourth class is composed of millisecond pulsars, whose
luminosity strongly depends on the predominant mechanism
generating X-ray quanta (residual accretion; collision of the pulsar
wind with the ambient medium or with matter lost by the companion;
thermal surface emission).

The study of weak X-ray sources is hampered by the fact that even in
the nearest GCs they are detected down to the present sensitivity
limit, and often remain unclassified as they are too dim to allow
for an estimate of the hardness ratio. Most of them are likely
active binaries or cataclysmic variables, but some may be quiescent
systems with neutron stars or even black holes. Optical counterparts
of these systems are often weak and hard to identify in the crowded
GC environment \citep[e.g.,][]{ver08}. So far, the X-ray data have
allowed us to identify a few candidate qLMXBs, but their nature has
not been confirmed by optical spectroscopy \citep{gui09}.

We decided to follow an entirely different approach. Instead of
looking for optical counterparts of X-ray sources, we select a sample of
candidate systems for degenerate binaries from short-period, low-amplitude
optical variables catalogued in existing surveys, and measure their radial
velocities. Our targets have nearly sinusoidal light curves with $V$-band
amplitudes smaller than 0.35~mag and periods shorter than $\sim$1.3~days.
This type of variability is common in all classes of close degenerate
binaries; in particular, it is observed in optical counterparts of several
Galactic X-ray novae which almost certainly harbor BHs \citep{rem06}.
The optical modulation is induced mostly by the
ellipsoidal effect from the nondegenerate component, with an amplitude
depending mainly on the Roche lobe filling factor and the inclination
of the orbit. Close degenerate binaries can be unambiguously identified as
single-line systems with large orbital velocities ($K >150$~km~s$^{-1}$).
Among them, those harboring a black hole are distinguished by a mass function
$f_m = (m\sin i)^3/(m_{bin})^2 >2M_{\odot}$, where $m_{bin}$ and $m$ stand
for the total mass of the binary and the mass of one of its components
\citep{rem06}.

Our sample and observational data are introduced in Sect.~2.
A detailed analysis of the data is reported in Sect.~3,
and the results are discussed in Sect.~4.

\section {Observations and data reduction}

Optical counterparts of active X-ray sources in GCs have been found essentially everywhere in
color-magnitude diagrams: to the left of the main sequence, on the main sequence, to the right
of it, on subgiant, giant and horizontal branches, and also above the main-sequence turnoff
in the regions occupied by blue stragglers and EHB stars (e.g. Heinke et al. 2005; Servillat
et al. 2008). Thus, while preparing this study we felt it justified to pick photometrically
suitable targets without paying attention to their location with respect to the main sequence.
Since large-amplitude variables are more likely to be ordinary
contact binaries, we avoided objects with
full $V$-band amplitudes in excess of $\sim$0.35~mag. Suspected
pulsating variables were excluded; we also took care to exclude
systems with appreciable X-ray emission. Where it was possible, the
membership status of the target was established based on the data of
\cite{bel09} for $\omega$~Cen, and Anderson (priv. comm.) and \cite{str09}
for NGC 6397. In the remaining cases membership was decided {\bf\it a posteriori}
based on the systemic velocity with respect to the cluster.

\begin{figure}[t]
\centering
\includegraphics[width=0.5\textwidth]{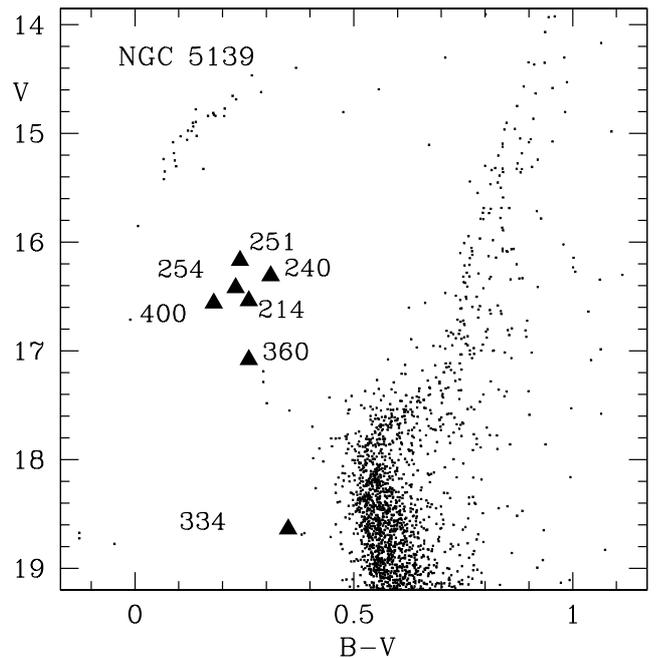}
\caption {Color-magnitude diagram of $\omega$~Cen based on CASE data
\citep{kal04} transformed to the BV system with the help of
standards listed by \cite{ste00}. Marked are the locations of the
investigated systems.}
\label{fig: omegacmd}
\end{figure}

\begin{figure}[h!]
\centering
\includegraphics[width=0.5\textwidth]{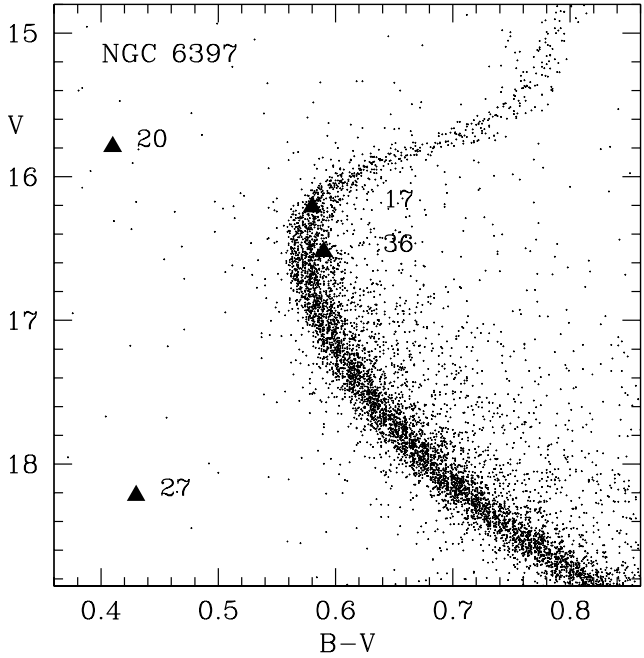}
\caption {Color-magnitude diagram of NGC 6397 \citep[after ][]{kal06}
with marked positions of the investigated systems.}
\label{fig: n6397cmd}
\end{figure}

Alltogether, we selected seven objects in $\omega$ Cen and four in NGC 6397.
In color-magnitude diagrams two of them are located to the left of the main
sequence, seven above the main-sequence turnoff, and two at the turnoff itself
(see Figs. \ref{fig: omegacmd} and \ref{fig: n6397cmd}).
A summary of their basic data is given in Table~\ref{tab: objects}
(the labels are the same as those introduced by \cite{kal04} for $\omega$~Cen
and \cite{kal06} for NGC 6397).

Our paper is based on photometric measurements of
\cite{kal04,kal06}, supplemented by those of \cite{wel07} and by
our unpublished data for NGC 6397 obtained on the du Pont telescope
at Las Campanas Observatory, Chile (a total of 771 $V$-band frames was acquired between
2009 June 20 and June 30, and reduced in the same way as in \cite{kal06}).
The newer data demonstrated that the observed light
variations were coherent, although in some cases the
light curve was changing from symmetric to asymmetric or vice versa.

Spectroscopic data were collected during the nights 20/21 and 21/22
of May 2009 with the MagE (Magellan Echellette) spectrograph attached to
the 6.5-m Magellan-Clay telescope at Las Campanas Observatory. The seeing
varied betweeen $0\farcs5$ and $0\farcs8$ on the first night, and betweeen
$0\farcs4$ and $0\farcs9$ on the second one. A $0\farcs85$ slit was used, providing
a resolution $R=4820$.

During the observations pairs of scientific spectra
taken for the same target were separated by an exposure of a thorium-argon
hollow-cathode lamp. The exposure times per spectrum ranged from 120~s
to 1080~s, depending on object brightness and observing conditions. After
bias and flat-field correction each pair of frames was combined into a
single frame, allowing for the rejection of cosmic ray hits. The
observations were reduced with the IRAF\footnote
{IRAF is distributed by the National Optical Astronomy Observatories,
 which are operated by the AURA, Inc., under cooperative agreement
 with the NSF.} ECHELLE package.

\begin{table*}[t]
\begin{center}
\caption{Basic information on the target objects.}
  {\small
   \begin{tabular}{lcccccccc}
    \hline
    Cluster & Star & $\alpha_{2000}$ & $\delta_{2000}$ & $p_m$* & $L_X$** & $V$ & $B-V$ & ref.$\dagger$ \\
      & & [h:m:s]  & [$\degr$:$\arcmin$:$\arcsec$] & & [$10^{30}$ erg~s$^{-1}$] & [mag] & [mag] & \\
    \hline
    $\omega$~Cen &  V214 & 13:27:21.82 & $-47$:37:19.0 &~~97 & $<$13& 16.55 & 0.25 & 1,2\\
    $\omega$~Cen &  V240 & 13:27:28.68 & $-47$:26:19.5 &~~99 &$<$1.2& 16.31 & 0.30 & 1,2\\
    $\omega$~Cen &  V251 & 13:27:28.02 & $-47$:26:43.7 &~~97 &$<$1.2& 16.17 & 0.25 & 1\\
    $\omega$~Cen &  V254 & 13:27:28.61 & $-47$:27:39.0 &~~99 &$<$1.2& 16.41 & 0.25 & 1,2\\
    $\omega$~Cen & NV334 & 13:27:23.39 & $-47$:29:11.9 &~~88 &$<$1.2& 18.63 & 0.35 & 1\\
    $\omega$~Cen & NV360 & 13:26:02.06 & $-47$:32:23.6 & 100 &$<$1.2& 17.09 & 0.25 & 1\\
    $\omega$~Cen & NV400 & 13:26:10.09 & $-47$:31:50.3 &~~99 &$<$1.2& 16.56 & 0.20 & 1\\
    NGC 6397     &  V17  & 17:40:43.78 & $-53$:41:16.2 & 100 &   0.4& 16.17 & 0.40 & 3,4\\
    NGC 6397     &  V20  & 17:40:41.66 & $-53$:40:33.1 & 100 &   1.6& 15.75 & 0.40 & 3,4\\
    NGC 6397     &  V27  & 17:41:13.80 & $-53$:41:14.1 &  -  &     -& 18.19 & 0.45 & 3,4\\
    NGC 6397     &  V36  & 17:40:44.10 & $-53$:42:11.3 &  -  &   0.4& 16.48 & 0.60 & 3,4\\
    \hline
   \label{tab: objects}
   \end{tabular}
  }
\end{center}
{\footnotesize * Membership probability: \cite{bel09} for V214-NV400;
 \cite{str09} and Anderson (2010, priv. comm.) for V17-V20; no data available
for V27 and V36.\\
 {**} X-ray luminosity: \cite{gen03} for V214; \cite{hag04} for V240-NV400;
 \cite{bog10} for V17-V20 and V36; no data available for V27.\\
 $\dagger$ Photometry: 1 - \cite{kal04}; 2 - \cite{wel07}; 3 - \cite{kal06};
 4 - our unpublished data collected in 2009.
}
\end{table*}

In order to determine the mass function of a given target it was
sufficient to take just a few spectra at different phases, and fit a
simple sinusoid to its velocity curve phased with the photometric
ephemeris. Altogether, 96 spectra were obtained (from 5 to 10 per
object). The useful range of the reduced spectra extended from
4050~\AA\ to 6800~\AA, where the spectra had
$13<\mathrm{S/N}<23$ (with a few cases of lower quality). Radial
velocities were measured in that range with the help of IRAF routines FXCOR and
XCSAO, using synthetic templates from the library compiled by \cite{mun05}.
The results were verified by applying the broadening
function formalism described by \cite{ruc02}. In most cases the
differences between various measurements were smaller than $\Delta v
= 10$ km s$^{-1}$, which may be regarded as a fair estimate of the
error.

\section {Analysis}

While our data are insufficient for a detailed analysis, an estimate of
the basic parameters of the observed systems is possible. In all cases
but one (V20 in NGC6307) single-line spectra are observed, indicating that to a first
approximation all the light we receive from the binary is produced by just
one of the components. Based on this assumption, we may set some
limits for acceptable configurations.

\subsection {Method}
\label{sect: method}
Our observables are: orbital period $P$, apparent $V$-band magnitude $V$, color index
$B-V$, amplitude of the light curve $\Delta V_{obs}$, and amplitude of the
velocity curve $K_{obs}$. Since the distance moduli to $\omega$~Cen and NGC~6397
are known, $V$ of a system belonging to the cluster is
equivalent to the observed bolometric magnitude of its visible component $M^o_{bol}$
("observed" means here "derived from observations; hence index "o").
We adopt $(m-M)_0 = 13.75$~mag and $E(B-V)$ = 0.115~mag for $\omega$~Cen
\citep[][ and references therein]{vil07}, and $(m-M)_0 = 12.03$~mag
and $E(B-V) = 0.183$~mag for NGC 6397 \citep[][ and references therein]{ric08}.
Assuming $A_V = 3.1\, E(B-V)$, the corresponding values of $A_V$ are 0.36~mag
and 0.57~mag, yielding a distance modulus $(m-M)_V$ of 14.11~mag for
$\omega$~Cen and 12.60~mag for NGC 6397.

The effective temperature of the visible component $T$ is known from the relation
between the color index $B-V$ and temperature derived for an appropriate
chemical composition. We adopt ${\rm [Fe/H]} = -1.8$ for $\omega$~Cen
\citep{sol09} and ${\rm [Fe/H]} = -2.0$ for NGC 6397 \citep{car09},
and we use color transformation tables provided by \cite{lej98}.
The parameters derived for our target stars from the observational data are
given in Table~\ref{tab: obspar}.

\begin{table*}[t]
\begin{center}
\caption{Parameters of the target objects derived from the observations.}
  {\small
   \begin{tabular}{lcclcrrr}
    \hline
    Cluster & Star & $P$ & $M^o_{bol}$ & $T$ & $K_{obs}$ & $f_m$ & $v_0$ \\
        & & [d] & [mag] & [K] & [km s$^{-1}$] & [$M_\odot$] & [km s$^{-1}$] \\
    \hline
    $\omega$~Cen &  V214 & 0.341806 & 2.44 & 8300 & 35    & 1.5$\times10^{-3}$ &   21\\
    $\omega$~Cen &  V240 & 0.331888 & 2.19 & 8000 & $<10$ & $<3.4\times10^{-5}$&   21\\
    $\omega$~Cen &  V251 & 0.922458 & 2.06 & 8300 & 35    & 4.1$\times10^{-3}$ &    6\\
    $\omega$~Cen &  V254 & 0.385071 & 2.30 & 8300 & 60    & 8.0$\times10^{-3}$ &   18\\
    $\omega$~Cen & NV334 & 0.257881 & 4.52 & 7300 & 35    & 1.1$\times10^{-3}$ & $-9$\\
    $\omega$~Cen & NV360 & 0.630821 & 2.98 & 8300 & 45    & 5.9$\times10^{-3}$ &$-17$\\
    $\omega$~Cen & NV400 & 0.636874 & 2.40 & 8800 & 20    & 5.3$\times10^{-4}$ & $-7$\\
    NGC 6397     &  V17  & 1.061316 & 3.35 & 6200 & 50    & 1.4$\times10^{-2}$ & $-2$\\
    NGC 6397     &  V20  & 0.861177 & 2.94 & 7200 & *     &                   *&    9\\
    NGC 6397     &  V27  & 0.556134 & 5.38 & 7500 & 25    & 9.0$\times10^{-4}$ & $-6$\\
    NGC 6397     &  V36  & 1.098569 & 3.65 & 6200 & 95    & 9.7$\times10^{-2}$ & $-1$\\
    \hline
   \label{tab: obspar}
   \end{tabular}
  }
\end{center}
{\footnotesize * Velocities of both components were measured, yielding
 $K_1 = 25$~km~s$^{-1}$ and $K_2 = 130$~km~s$^{-1}$. The corresponding mass functions
 are $f_{m1}=1.4\times10^{-3}$~$M_\odot$ and $f_{m2}=1.9\times10^{-1}$~$M_\odot$.}
\end{table*}

The last column in Table~\ref{tab: obspar} contains mass-center velocities
with respect to the cluster obtained by a simple sinusoid fit, with
heliocentric velocities of the clusters (232.3 km s$^{-1}$ for $\omega$~Cen
and 18.9~km~s$^{-1}$ for NGC 6397) taken from the catalogue compiled by \cite{har96}
and updated online.\footnote{http://www.physics.mcmaster.ca/Globular.html}
The velocities of V214, V240, and V254 may seem a bit largish for their distances
from the cluster center ($17\farcm0$, $14\farcm9$, and $14\farcm7$, respectively),
even though the published cluster tidal radius is $r_t = 44\farcm8$ \citep{tea95},
or $r_t = 57\farcm03$ according to the \citet{har96} catalog. That
these stars are well within the limits for cluster membership is confirmed
by Fig.~6 in \citet{sol09}, which shows the variation in radial velocity of
$\omega$~Cen stars as a function of distance from the cluster center.
Still, V240 turned out to be a doubtful case (see Sect.
\ref{sect: omegaCen}). As for V214 and V254, we checked that the best fits
described in Sect. \ref{sect: omegaCen} were not affected when $v_0$ was lowered,
respectively, to 10 and 6~km~s$^{-1}$. The
velocity of NV360, although also rather large, is entirely consistent with the
distance of this system from the center of $\omega$~Cen ($3\farcm6$), at which
the relative velocities of cluster members reach $\pm 35$~km~s$^{-1}$
\citep{sol09}.

With $T$ being fixed, the task is to find the orbital separation $a$,
inclination of the orbit $i$, and masses of the components $m_1$ and $m_2$.
Unless indicated otherwise, we assume that the system is semi-detached, with
the visible component filling its Roche lobe.
It must be stressed here that this particular model of the binary {\em is not
related to the physical structure of our targets} (in fact, in some cases it might imply
a mass transfer unstable on a dynamical scale).
A semi-detached system is simply the most compact one among those with $M^c_{bol}=M^o_{bol}$.
Focusing on it, we minimize the orbital separation, so that the corresponding
``semi-detached'' masses $m_1^{sd}$ and $m_2^{sd}$ are the smallest allowable given $P$
and $q$. The lower limits of the actual masses of the components are obtained by finding
``the best'' mass ratio $q_b$, defined as the one for which $\Delta V_c/\Delta V_{obs}=1$
(see Sect. \ref{sect:errors} for further discussion).
Our analysis involves the following steps:
\begin{enumerate}
 \item Specify if the visible component is primary or secondary.
 \item Set the mass ratio $q\equiv m_2/m_1<1$ (we define the primary
  as the more massive component, not necessarily being the more luminous one).
 \item Given the orbital period $P$, adjust the separation $a$ so that
  the calculated bolometric magnitude $M^c_{bol}$ of the visible component
  be equal to $M^o_{bol}$.
 \item Adjust $i$ and $v_0$ so that the calculated velocity curve fits the
  observed one, and find the ratio of the computed amplitude of the light
  curve ($\Delta V_c$) to the observed amplitude ($\Delta V_{obs}$).
 \item Repeat steps (2)-(4) for several $q$ values.
 \item If in step (1) the visible component was specified as the primary,
  specify it as the secondary (and {\it vice-versa}). Repeat steps (2)-(4) for
  several values of $q$.
\end{enumerate}
The calculations in steps (3) and (4) are performed using the PHOEBE
interface \citep{prs05} to the Wilson-Devinney code \citep{wil71}.

In principle, using a relation between
$V$, $B-V$ and stellar angular diameter $\theta$ \citep{ker04}
would be more straightforward and simpler than calculating
$M^o_{bol}$ and $T$, and comparing $M^o_{bol}$ to $M^c_{bol}$ (note
that $T$ is needed for PHOEBE to derive $M^c_{bol}$). Based on such
a relation, we could directly fit the ``observed'' radius of the
visible component $R^o\equiv\theta D$ (where $D$ is the distance to
the cluster) to the calculated average radius of the Roche lobe.
Unfortunately, fits of \cite{ker04} proved to be
unreliable when extrapolated to low metallicity and small $\theta$.
This is not surprising, as these authors explicitly warn
about the nonlinearity of the relation involving $V$ and $B-V$.

\subsection {Discussion of errors}
\label{sect:errors}

The values of our key input parameters, i.e. temperatures and
bolometric luminosities, are not known accurately, and their errors
are difficult to estimate without engaging in a lengthly (and likely
ambiguous) discussion of involved factors. In order to verify how
these uncertainties influence the output we varied $T$ and
$M^o_{bol}$ by small amounts and observed the corresponding
variations in $m_1$ and $m_2$. We found that decreasing $T$ by 100~K
causes the masses to grow by $7-10$\%, while an increase in
$M^o_{bol}$ by $-0.1$~mag makes them larger by $10-15$\%.

At a first glance, focusing on semi-detached systems seems too
restrictive. To see it clearly, suppose that the visible component
is entirely contained within its Roche lobe. Then in order to
recover $M^o_{bol}$ we would have to increase the orbital separation $a$.
Since the period $P$ is
fixed, the masses would have to increase, too. To keep the observed
velocity amplitude constant, the inclination $i$ would have to decrease. For that
reason, and also because the visible component would now be less
deformed, $\Delta V_c$ would also decrease. This way the range of $q$ for
which the ratio $\Delta V_c/\Delta V_{obs}\approx1$ could be extended onto fits
which for semi-detached systems yield $\Delta V_c/\Delta V_{obs}>1$. In principle,
it is even possible that $m_1$ and $m_2$ derived for some mass ratio $q\ne q_b$
from detached fits could be smaller than $m_1^{sd}(q_b)$ and
$m_2^{sd}(q_b)$.
However, upon verifying this possibility we found that in most
cases the detached fit results in either too low an amplitude
for the light-curve or masses larger than those obtained for $q_b$, so that
the extension of the $q$-range is marginal (if any). We conclude
that the errors in the mass limits caused by neglecting detached
configurations are smaller than those related to bolometric
corrections or the color-temperature scale.

Another possible source of
errors is the assumption that the whole light output of the system
originates in the component responsible for the observed spectrum.
The components of the only system in our sample in which two sets of
velocities could be measured differ in brightness by $\Delta
M\approx 1$~mag. Assuming that in the remaining systems $\Delta M =
2$~mag causes their brighter components to be dimmer by 0.16~mag.
The corresponding decrease in masses, estimated from the general
relation between mass and $M^o_{bol}$ mentioned earlier, amounts
to $15-20$\%.

\subsection {Results} \label{sect: Results}

\begin{figure}[h!]
\centering
\includegraphics[width=0.5\textwidth]{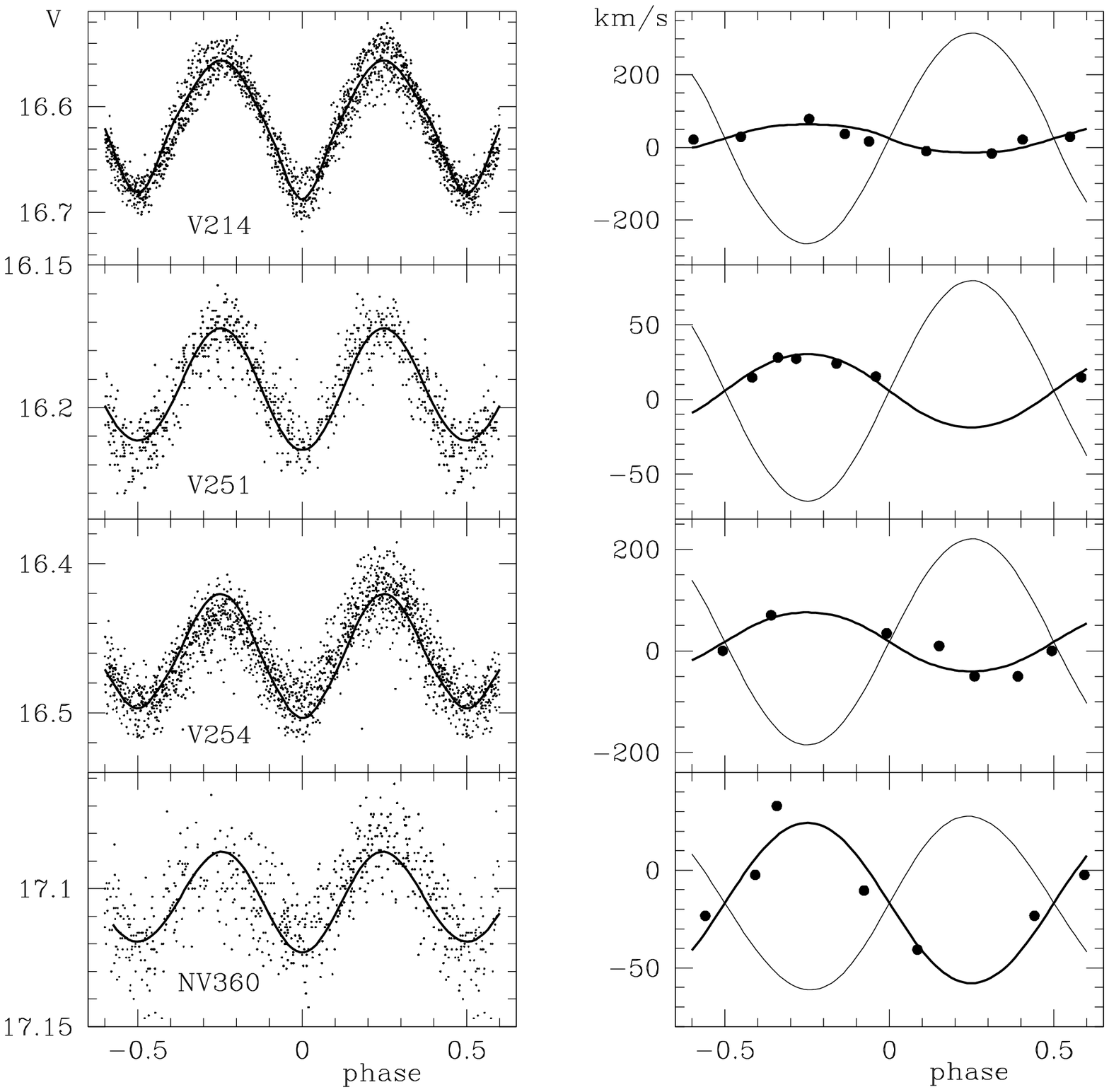}
\caption{Light and velocity curves of blue stragglers.
Parameters of the plotted fits are given in Tables \ref{tab: V214c},
\ref{tab: V251p}, \ref{tab: V254p}, and \ref{tab: NV360p}, in the rows
indicated with an asterisk.}
\label{fig: bs1}
\end{figure}

\begin{figure}
\centering
\includegraphics[width=0.5\textwidth]{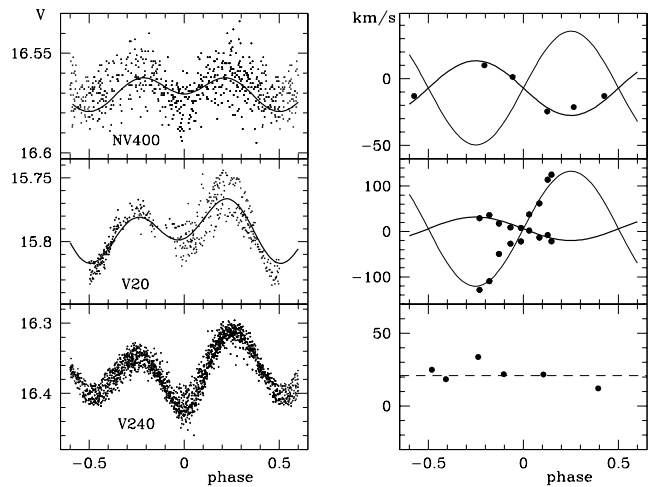}
\caption {Light and velocity curves of blue stragglers~-- continuation of
Fig. \ref{fig: bs1}. A single spot 2\% hotter than the photosphere was
placed on the primary of V20 in order to qualitatively account for the
asymmetry of the light curve (just for illustrative purposes; it was
not included while analyzing the data). Parameters of the fits for
NV400 and V20 are given in Tables \ref{tab: NV400p} and \ref{tab: V20}
in the rows indicated with an asterisk. No fit was possible for V240,
because the velocity ampliutude turned out tobe smaller than the
observational errors. The light curve, however, is coherent and
remained stable between 1999 and 2003.}
\label{fig: bs2}
\end{figure}

\begin{figure}
\centering
\includegraphics[width=0.5\textwidth]{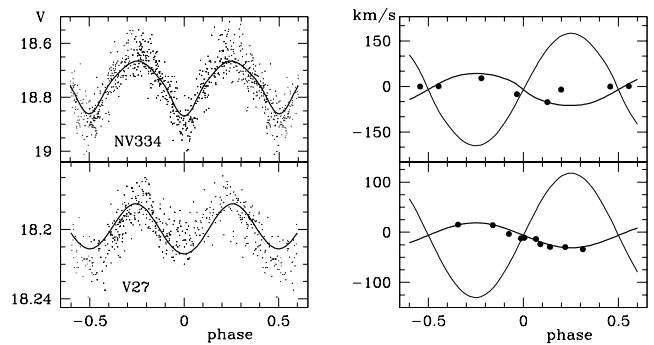}
\caption {Light and velocity curves of subdwarfs. Parameters of the
plotted fits are given in Tables \ref{tab: NV334p} and \ref{tab: V27}
in the rows indicated with an asterisk.}
\label{fig: sd}
\end{figure}

Quantitative results of the analysis are shown in Tables 3-13, and
discussed in Sect. \ref{sect: Results}. The best fits are indicated
with an asterisk in the last column of each table, and the
corresponding computed light and velocity curves are displayed in
Figs. \ref{fig: bs1}, \ref{fig: bs2}, \ref{fig: sd} and \ref{fig:
tms}, along with the observational data.

\subsubsection{$\omega$ Centauri} \label{sect: omegaCen}

\noindent{\bf V214}.
Assuming that the visible component is the secondary results in computed
light-curve amplitudes consistently much too low. Assuming that it is the
primary makes the difference between $\Delta V_{obs}$ and $\Delta
V_c$ smaller, but still unacceptable.

In order to make $\Delta V_c$ larger, we are forced to look for solutions in
which both components contribute to the observed ellipticity effect.
Based on the simplest possible assumption
that the system is in contact, we apply the following procedure: i) set $q$;
ii) adjust $a$ so that the combined $M_{bol}$ of the system is 2.44~mag; iii) adjust
$i$ so that the computed velocity curve agrees with the observed one; iv) check
if the computed light curve agrees with the observed one.

\begin{table}[h!]
\centering
 \caption{ \label{tab: V214c} Derived parameters for V214 }
 \begin{tabular}{cccccc}
  \hline
  $q$ & $a$ & $i$ &$m_1$ & $m_2$ &  $\Delta V_c/\Delta V_{obs}$ \\
      & [$R_\odot$] & [$\degr$] & [$M_\odot$] & [$M_\odot$] & \\
  \hline
   0.10 & 2.45 & 62 & 1.54 & 0.15 & 1.1  \\
   0.13 & 2.55 & 58 & 1.69 & 0.22 & 1.0* \\
   0.15 & 2.60 & 47 & 1.76 & 0.26 & 0.6  \\
  \hline
 \end{tabular}
\end{table}

The fits are
shown in Table~\ref{tab: V214c}, and we conclude that the available data favor a
system with a rather large mass ratio, which must have undergone significant
mass transfer (and may still be transferring mass at a low rate). Obviously,
the sum $m=m_1+m_2$ should not exceed two turnoff masses of the youngest
population of $\omega$~Cen, i.e. 1.84 $M_\odot$ \citep{nor04}. Given the rather
large errors in the mass estimates due to uncertainties in $T$ and $M^o_{bol}$, one
may accept that this requirement is fulfilled by at least the first two fits in Table
\ref{tab: V214c}.\\

\noindent{\bf V251}.
Like in the previous cases, assuming that the spectrum originates in the
secondary we find that for all $q$ values $\Delta V_c$ is much smaller than
$\Delta V_{obs}$. Assigning the spectrum to the primary leads
to the results displayed in Table~\ref{tab: V251p}. As the total mass
is markedly smaller than 1.84 $M_\odot$, we seem to have a system
which not only underwent mass transfer, but also lost a significant
amount of mass.\\

\begin{table}[h!]
\centering
%\vspace{-0.6 cm}
 \caption{ \label{tab: V251p} Derived parameters for V251 }
 \begin{tabular}{cccccc}
  \hline
  $q$ & $a$ & $i$ &$m_1$ & $m_2$ &  $\Delta V_c/\Delta V_{obs}$  \\
      & [$R_\odot$] & [$\degr$] & [$M_\odot$] & [$M_\odot$] & \\
  \hline
   0.20 & 3.10 & 33 & 0.39 & 0.08 & 3.2 \\
   0.35 & 3.44 & 32 & 0.48 & 0.17 & 1.0*\\
   0.50 & 3.67 & 22 & 0.52 & 0.26 & 0.65 \\
  \hline
 \end{tabular}
\end{table}

\noindent{\bf V254}. Assuming that the spectrum originates from the
secondary leads once more to amplitude ratios $\Delta V_c/\Delta V_{obs}$ which
for all mass ratios are much smaller than unity. Upon adopting that the source
of the spectrum is the primary we get the results displayed in
Table~\ref{tab: V254p}, which suggests that V254 is similar to V214.\\

\begin{table}[h!]
\centering
%\vspace{-0.6 cm}
 \caption{ \label{tab: V254p} Derived parameters for V254 }
 \begin{tabular}{cccccc}
  \hline
  $q$ & $a$ & $i$ &$m_1$ & $m_2$ &  $\Delta V_c/\Delta V_{obs}$  \\
      & [$R_\odot$] & [$\degr$] & [$M_\odot$] & [$M_\odot$] & \\
  \hline
   0.2 & 2.77 & 72 & 1.61 & 0.32 & 1.7  \\
   0.3 & 2.98 & 43 & 1.84 & 0.55 & 1.0* \\
   0.5 & 3.29 & 24 & 2.15 & 1.07 & 0.6  \\
  \hline
 \end{tabular}
\end{table}

\noindent{\bf NV334}. Because of the low brightness of the system
($V=18.83$~mag) we were able to obtain only a very rough estimate
of the velocity amplitude. We therefore modifed our analysis:
we estimated the inclination based on the light curve, and compared
the calculated velocity amplitude $K_c$ to the observed one. Assuming
that the spectrum originates in the secondary leads to $K_c$ being much
larger than $K_{obs}$, regardless of the value of $q$. The same analysis
applied to the configuration with the primary generating the spectrum
produces results shown in Table
\ref{tab: NV334p}, which seem to indicate that NV334 is similar to
V251. However, the similarity is superficial: the effective
temperatures of the two systems differ by $\sim4000$ K, causing the
cooler and dimmer NV334 to occupy an entirely different position in
the color-magnitude diagram of $\omega$~Cen (see Fig. \ref{fig:
omegacmd}).\\

\begin{table}[h!]
\centering
%\vspace{-0.6 cm}
 \caption{ \label{tab: NV334p} Derived parameters for NV334 }
 \begin{tabular}{cccccc}
  \hline
  $q$ & $a$ & $i$ &$m_1$ & $m_2$ &  $K_c/K_{obs}$  \\
      & [$R_\odot$] & [$\degr$] & [$M_\odot$] & [$M_\odot$] & \\
  \hline
   0.1 & 1.22 & 76 & 0.33 & 0.03 & 0.4 \\
   0.3 & 1.45 & 65 & 0.47 & 0.14 & 1.2*  \\
   0.5 & 1.60 & 63 & 0.55 & 0.27 & 1.8  \\
  \hline
 \end{tabular}
\end{table}

\noindent{\bf NV360}. Our standard analysis fails for this system
(iterations of the inclination $i$ based on the velocity curve are either converging
extremely slowly or entirely diverging), so that we again have to
apply a modified version based on the light curve. As for the
case of NV334, assuming that the spectrum originates in the
secondary leads to $K_c$ being much larger than $K_{obs}$, regardless of
the value of $q$. Systems with the primary being responsible for the
spectrum fare much better (see Table~\ref{tab: NV360p}), and we
conclude that NV360, unlike NV334, is truly similar to NV251.\\

\begin{table}[h!]
\centering
%\vspace{-0.6 cm}
 \caption{ \label{tab: NV360p} Derived parameters for NV360 }
 \begin{tabular}{cccccc}
  \hline
  $q$ & $a$ & $i$ &$m_1$ & $m_2$ & $K_c/K_{obs}$  \\
      & [$R_\odot$] & [$\degr$] & [$M_\odot$] & [$M_\odot$] & \\
  \hline
   0.5 & 2.52 & 26 & 0.36 & 0.18 & 0.7  \\
   0.8 & 2.79 & 25 & 0.41 & 0.32 & 1.0*  \\
   0.9 & 2.86 & 25 & 0.42 & 0.37 & 1.1  \\
  \hline
 \end{tabular}
\end{table}

\noindent{\bf NV400}. When the secondary is assumed to generate the
spectrum, the calculated light-curve amplitude is for all values of
$q$ much higher than the observed one. Assuming that the spectrum
originates in the primary we get the results displayed in Table
\ref{tab: NV400p}. Apparently, NV400 is yet another system similar
to V251.\\

\begin{table}[h!]
\centering
%\vspace{-0.6 cm}
 \caption{ \label{tab: NV400p} Derived parameters for NV400 }
 \begin{tabular}{cccccc}
  \hline
  $q$ & $a$ & $i$ &$m_1$ & $m_2$ &  $\Delta V_c/\Delta V_{obs}$  \\
      & [$R_\odot$] & [$\degr$] & [$M_\odot$] & [$M_\odot$] & \\
  \hline
   0.3 & 2.65 & 28 & 0.48 & 0.14 & 1.3  \\
   0.5 & 2.93 & 13 & 0.56 & 0.28 & 0.8* \\
   0.8 & 3.24 &  9 & 0.63 & 0.50 & 0.2  \\
  \hline
 \end{tabular}
\end{table}

\noindent{\bf V240}.
Assuming that the spectrum originates from the primary we obtain the results
shown in Table~\ref{tab: V240p}: computed
light-curve amplitudes are far too low (if the secondary were the source of
the spectrum they would be even lower). Moreover, the total mass of the system
is unacceptably high (as we argued in  the case of V214, it should not exceed
1.84 $M_\odot$). In other words, the color and apparent magnitude of V240 are
incompatible with its observed velocity amplitude, suggesting that this system
does not belong to $\omega$~Cen. Another possibility is that the light of V240
is dominated by a tertiary component, and the recorded spectra are unrelated to
the photometric binary.\\

\begin{table}[h!]
\centering
%\vspace{-0.9 cm}
 \caption{ \label{tab: V240p} Derived parameters for V240 }
 \begin{tabular}{cccccc}
  \hline
  $q$ & $a$ & $i$ &$m_1$ & $m_2$ &  $\Delta V_c/\Delta V_{obs}$ \\
      & [$R_\odot$] & [$\degr$] & [$M_\odot$] & [$M_\odot$] & \\
  \hline
   0.07 & 2.83 & 16 & 2.59 & 0.18 & 0.1 \\
   0.1  & 2.96 & 11 & 2.88 & 0.29 & 0.05 \\
  \hline
 \end{tabular}
\end{table}

\begin{figure}
\centering
\includegraphics[width=0.5\textwidth]{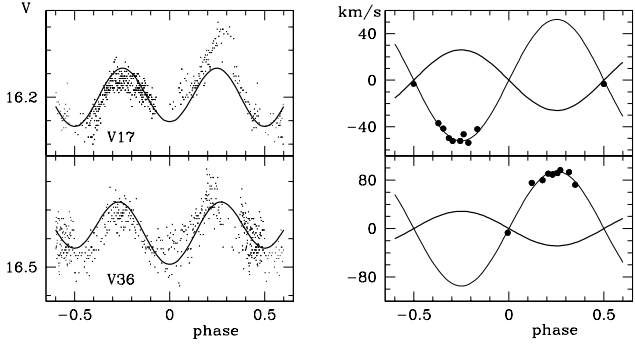}
\caption {Light and velocity curves of systems at the main sequence
turnoff. Small ticks in light-curve plots are separated by 0.01~mag.
Parameters of the plotted fits are given in Tables \ref{tab: V17}
and \ref{tab: V36} in the rows indicated with an asterisk.}
\label{fig: tms}
\end{figure}

\subsubsection{NGC 6397} \label{sect: NGC6397}

\noindent{\bf V17}.
The visible component must be the secondary (an alternative assumption enforces
inclinations at which, regardless of the value of $q$, the ratio $\Delta V_c/\Delta V_{obs}$
is much lower than unity). Our standard analysis produced the results shown in
Table~\ref{tab: V17}.

The fact that V17 is situated at the turnoff of the main sequence
prompted us to consider an exceptional case of a detached binary
with $m_2$ equal to the turnoff mass, which in NGC 6397 is close
to 0.8 $M_\odot$ \citep{kal08}. Keeping a conservative
approach we set $m_2=0.75$ $M_\odot$, and, as expected
based on Table~\ref{tab: V17}, we found that the amplitudes of all
calculated light curves were too low. Our results indicate that
the primary, although dimmer than the secondary, is rather massive,
with a mass possibly exceeding 1 $M_\odot$.

\begin{table}[h!]
\centering
%\vspace{-0.9 cm}
 \caption{ \label{tab: V17} Derived parameters for V17 }
 \begin{tabular}{cccccc}
  \hline
  $q$ & $a$ & $i$ & $m_1$ & $m_2$ & $\Delta V_c/\Delta V_{obs}$  \\
      & [$R_\odot$] & [$\degr$] & [$M_\odot$] & [$M_\odot$] & \\
  \hline
   0.3 & 6.00 & 14 & 1.99 & 0.60 &  0.7 \\
   0.4 & 5.56 & 16 & 1.47 & 0.59 &  0.9 \\
   0.5 & 5.25 & 19 & 1.15 & 0.58 &  1.1* \\
   0.6 & 5.02 & 21 & 0.95 & 0.57 &  1.5 \\
  \hline
 \end{tabular}
\end{table}

\noindent{\bf V20}.
This is the only system in our sample for which it was possible to measure
the velocities of both components. The measurements were performed using
the TODCOR method developed by \cite{zuc94}, which also calculates
the ratio of component fluxes $\alpha\equiv f_1/f_2$. Based on all available
spectra of V20 (there were 9 of them), we obtained $\alpha=2.47\pm0.02$ which,
combined with $M^o_{bol} = 2.94$~mag, yields $M^o_{bol1}=3.31$~mag and
$M^o_{bol2}=4.29$~mag.

Since the mass ratio is small ($q\approx0.2$), the secondary must be rather
oversized for its mass. We assume that it fills its Roche lobe, and we apply
the following nonstandard procedure: i) set $i$; ii) iterate $a$ and $q$ to
fit the velocity curves of both components); iii) adjust the size of the
primary to get $M^c_{bol1}=3.31$~mag; adjust the temperature of the secondary
$T_2$ to get $M^c_{bol2}=4.29$~mag (note that this is the minimum possible
temperature). The best fit among those displayed in Table~\ref{tab: V20}
indicates that V20 is another system similar to V251.\\

\begin{table}[h!]
\centering
%\vspace{-0.9 cm}
 \caption{ \label{tab: V20} Derived parameters for V20 }
 \begin{tabular}{ccccccc}
  \hline
  $i$ & $a$ & $q$ & $m_1$ & $m_2$ & $\Delta V_c/\Delta V_{obs}$ \\
      & [$R_\odot$] & [$\degr$] & [$M_\odot$] & [$M_\odot$] & \\
  \hline
   30 & 5.28 & 0.20 & 2.23 & 0.45 & 0.4 \\
   40 & 4.09 & 0.20 & 1.04 & 0.20 & 0.8 \\
   45 & 3.72 & 0.20 & 0.78 & 0.15 & 1.0*\\
   50 & 3.43 & 0.20 & 0.61 & 0.12 & 1.2 \\
  \hline
 \end{tabular}
\end{table}

\noindent{\bf V27}. Upon applying the standard analysis we found
that, regardless of whether the primary or the secondary produces the
spectrum, for a given value of $q$ either there is no inclination
for which the calculated amplitude matches the observed one or the
masses of both stars are unacceptably low ($m_1\le0.05 \, M_\odot$;
$m_2\le0.04 \, M_\odot$). One can make them higher by assuming that
the system is detached instead of semi-detached; however, in such a case
it is impossible to find firm
lower mass limits.

\begin{table}[h!]
\centering
%\vspace{-0.9 cm}
 \caption{ \label{tab: V27} Derived parameters for V27 }
 \begin{tabular}{ccccccc}
  \hline
  $q$ & $a$ & $i$ & $m_1$ & $m_2$ & $T_2$ &  $\Delta V_c/\Delta V_{obs}$ \\
      & [$R_\odot$] & [$\degr$] & [$M_\odot$] & [$M_\odot$] & \\
  \hline
   0.2 & 2.00 & 55 & 0.29 & 0.06 & 6500 & 1.0*   \\
   0.2 & 2.50 & 40 & 0.57 & 0.11 & 6000 & 0.9 \\
   0.2 & 3.00 & 33 & 0.98 & 0.20 & 6000 & 0.3   \\
   0.3 & 1.80 & 43 & 0.20 & 0.06 & 6000 & 0.9 \\
   0.3 & 2.00 & 36 & 0.27 & 0.08 & 6000 & 0.7 \\
   0.3 & 2.50 & 28 & 0.52 & 0.17 & 6000 & 0.2   \\
  \hline
 \end{tabular}
\end{table}

To obtain a sample of detached solutions we modified our standard procedure in the
following way: i) set $q$ and $a$; ii) adjust $i$ so that the calculated velocity
curve agrees with the observed one; iii) adjust the size of the primary to get
$M_{bol1}$ = 6.0~mag; iv) adjust the size (and, if necessary, the temperature) of
the secondary to get $M_{bol2}$ = 6.3~mag (the latter two values are of course arbitrary,
the only constraints being that the secondary is not much dimmer than the primary,
and the combined bolometric magnitude of the system is 5.38~mag), v) check if the
calculated light curve agrees with the observed one.

If one assumes that the spectrum is generated by the secondary, then the
inclination consistent with the velocity curve is much too low to reproduce
the light curve (except for systems with $q\gtrsim0.75$, but in those cases the
masses of the components are smaller than 0.05 $M_\odot$).
The only reasonable solutions we were able to find for the detached configuration
are based on the assumption that the primary generates the observed spectrum,
while the secondary is mainly responsible for the observed ellipticity effect.
The results shown in Table~\ref{tab: V27} seem to favor a configuration with small
and rather discrepant masses.\\

\noindent{\bf V36}. The visible component must be the secondary; the
alternative assumption enforces inclinations such that the calculated
amplitude of the light curve is much lower
than the observed one for all values
of $q$. However, assuming that the secondary fills
its Roche lobe is untenable, as our standard analysis indicates that
in such a case the calculated amplitudes of the light curve are too
high. Thus, we are forced to adopt a detached configuration. Since
V36 is located at the turnoff (only 0.3~mag below V17; see Fig.~\ref{fig: n6397cmd}),
we assume that $m_2 = 0.75 \, M_\odot$ and apply the following four-step
procedure: i) set $q$; ii) adjust $a$ to get $m_2 = 0.75 \, M_\odot$; iii)~adjust
$i$ so that the calculated velocity curve agrees with the
observed one; iv) adjust the size of the secondary to get $M_{bol2}
= 3.65$~mag. The results shown in Table~\ref{tab: V36} indicate that V36
is similar to V17, with the primary's mass possibly even higher than 1.5~$M_\odot$.

\begin{table}[h!]
\centering
%\vspace{-0.9 cm}
 \caption{ \label{tab: V36} Derived parameters for V36 }
 \begin{tabular}{ccccccc}
  \hline
  $q$ & $a$ & $i$ & $m_2$ & $m_1$ &  $\Delta V_c/\Delta V_{obs}$ \\
      & [$R_\odot$] & [$\degr$] & [$M_\odot$] & [$M_\odot$] & \\
  \hline
   0.2 & 7.40 & 20 & 0.75 & 3.77 & 0.9   \\
   0.3 & 6.62 & 24 & 0.75 & 2.49 & 1.0*  \\
   0.4 & 6.17 & 28 & 0.75 & 1.87 & 1.3   \\
   0.5 & 5.86 & 32 & 0.75 & 1.50 & 2.1 \\
  \hline
 \end{tabular}
\end{table}

\section {Discussion and conclusions}

10 out of 11 objects in our sample exhibit radial-velocity
variations indicating their binary nature (the only exception is
V240 in $\omega$ Cen). In that sense, the sample was well chosen.
Regarding the principal aim of the present survey we have been less
successful - no clear-cut evidence for high-mass degenerate
components was found. However, while 8 systems proved to be more or
less ordinary binaries, the remainig two (V17 and V36 at the turnoff
of NGC 6397) cearly deserve further scrutiny. First, they are the
only systems in which the brighter component is the less massive
secondary. Second, the masses of their dim primaries may be
significantly larger than 1 $M_\odot$ (in V36, even larger than  2
$M_\odot$ if the secondary is indeed a turnoff-mass star). Third,
they are weak X-ray sources. Based on Chandra observations,
\cite{bog10} classify them as active binaries (AB), i.e. systems
composed of main-sequence or subgiant stars whose weak X-ray
emission originates from magnetic activity. Our results rule out the
possibility that they are composed of pristine stars which have not
undergone any mass transfer episode, as in such a case the more
massive component would also have to be the brighter one. This is
not to say \cite{bog10} are wrong: short-period degenerate binaries
in quiescence may easily ``masquerade'' as AB systems, since they
are most likely synchronized and their visible components spin fast
enough for the stellar dynamo to be highly efficient.

According to the most optimistic interpretation of our results, V17 and V36 may contain
a neutron star (the dim primary of V36 may even be a black hole). Such systems can indeed
be expected, as population synthesis calculations indicate that the formation of a
degenerate binary in which a neutron star or a black hole is accompanied by a main-sequence
star or a subgiant is nothing unusual in globular clusters. Although the rate at which they
form is not high ($<2.5$ systems/Gyr), they should be transient all the time, and therefore
more likely seen as a qLMXBs rather than as bright LMXBs \citep{iva08}. Moreover, NGC 6397
belongs to core-collapse clusters for which the formation rates of \cite{iva08} are just
lower limits. Another (admittedly, not too strong) support for the optimism comes from
the detection of a possible qLMXB near the turnoff of M30 \citep{lug06}. Main-sequence
secondaries similar to those in V17 and V36 are also found in galactic X-ray binaries
whose primary components are black hole candidates; e.g. V1033 Sco, GRS 1739--278 or V821
Ara (Ziolkowski, priv. comm).

The fact that all our blue stragglers except V240 are binary
supports the currently leading hypothesis concerning the nature of
these objects, namely that they result rather from an extensive mass
exchange between the component stars than from stellar collisions
\citep{knig09}. In all cases the brighter component is the primary
which according to the mass-transfer scenario must have acquired
significant amounts of hydrogen-rich material from the envelope of
the originally more massive secondary. In fact, our results seem to
indicate that blue stragglers are Algol-like systems in which the
original mass ratio has been reversed, causing the mass transfer to
effectively stop. Our stragglers may be similar to the well-studied
star V228 in 47 Tuc \citep{kal07}, however we cannot exclude the
possibility that in some of them the primary is already at the
beginning of the subgiant branch and has become large enough to
approach its Roche lobe.

V214 and V254 in $\omega$ Cen have primaries with masses likely
exceeding 1~$M_\odot$ (maybe even 1.5~$M_\odot$), and their total
masses approach 2~$M_\odot$. Systems that massive must have
originated from binaries subject to conservative or nearly
conservative mass transfer. They also have large amplitudes of light
variations (0.12,~mag and 0.10~mag, respectively), in our sample
rivaled only by that of NV334 ($\sim$0.3 mag); however a significant
contribution to this effect must originate from relatively large
inclinations found for all three systems. The remaining blue
stragglers are significantly less massive~-- their {\it total}
masses are lower than 1~$M_\odot$. As far as we know, our results
are the first to indicate such a broad (and possibly bimodal) mass
distribution. If this effect is real, it probably reflects
differences in the efficiency of mass transfer and/or mass loss. We
note in passing that some mass may still be flowing between the
components of V20, as this system exhibits an asymmetric and
variable light curve, and is a weak X-ray source. Another
possibility to account for the apparent large width of the mass
distribution is related to the peculiarities of the chemical
composition of $\omega$~Cen, in which strongly He-enriched
subpopulations are suggested to exist, with $Y$ reaching up to
$\sim0.4$ (e.g., \citeauthor{nor04} \citeyear{nor04};
\citeauthor{pio05} \citeyear{pio05}; but see also
\citeauthor{mcea10} \citeyear{mcea10} for a recent review and some
caveats).Our fits are based on the standard value $Y=0.24$, and for
a given $B-V$ color they can overestimate the temperature of
enriched stars by up to a few hundred Kelvin \citep{dot08}. If this
is what happened, then our standard analysis caused the He-rich
systems to artificially shrink both in size and mass. Such
hypothesis would also explain why stragglers with signficantly
different ``apparent'' masses are able to form a relatively tight
group on the H-R diagram.

The subdwarfs NV334 and V27 are very similar to each other in that they
have large mass ratios and low-mass primaries with practically the same
effective temperatures. Our estimate of the primary's mass in NV334
is consistent with the canonical hot subdwarf mass of 0.47 $M_\odot$,
while the estimated mass of the secondary falls near the peak of the
mass distribution of the unseen companions to field hot subdwarf stars
\citep{gei09}. On the other hand, and as pointed out by \citet{mbea08},
the close binary fraction among subdwarf stars appears to be much
lower in GCs than in the field, which may point to different formation
mechanisms for many of the GC subdwarfs.
In V27 the estimated masses are significantly smaller
(see Table~\ref{tab: V27}); note however that they are also less
reliable due to problems discussed in Section \ref{sect: NGC6397},
and that the solution with $m_1 = 0.57$~$M_\odot$ and $m_2 = 0.11$~$M_\odot$
is also acceptable. Taken at face value, the starred parameters in Table
\ref{tab: V27} would indicate that V27 is very similar to HS 2231+2241, an
HW Vir-type system with a companion at or below the very low mass limit
for M-dwarfs. The $\sim0.26$~$M_\odot$ primary of HS 2231+2241 is a
non-helium burning, post-RGB star \citep{ost08}. Another example
of such a type of star is the 0.24~$M_\odot$ component of HD 188112 \citep{heb03}.
The primaries of NV334 and V27 are much cooler ($T\sim7500$ K
compared to $T = 28400$ K and $T = 20500$ K, respectively, for HS 2231+2241
and HD 188112), but this difference may result from their more advanced
evolutionary stage.

Our sample is too small for a meaningful statistical analysis; nevertheless the
above-discussed results indicate that further research on ellipsoidal binaries in globular
clusters is a wortwhile task, and suggest its most promising direction: it seems that future
sampling of the photometric variables should be biased towards objects located on or to the
right of the main sequence, including those identified as weak X-ray sources. Obviously, for
purely economical reasons (exposure times!) the targets sould not fall much below the turnoff
point. Such an approach is tedious, but at the same time it is the only one on which a
fair census of degenerate binaries in globular clusters can be based.

\begin{acknowledgements}
We are very grateful to Jay Anderson for providing HST proper-motion
data for NGC 6397 and to Janusz Ziolkowski for a primer on black hole candidates in
galastic X-ray binaries. Research of JK, PP and WP is supported by
the grant MISTRZ from the Foundation for the Polish Science
and by the grants N~N203 379936 and N~N203 301335
from the Polish Ministry of Science and Higher Education.
Support for MC and CC is provided by MIDEPLAN's Programa Iniciativa
Cient\'{i}fica Milenio through grant P07-021-F, awarded to The Milky
Way Millennium Nucleus; by Proyecto Basal PFB-06/2007; by FONDAP
Centro de Astrof\'{i}sica 15010003; and by Proyecto FONDECYT
Regular \#1071002. IBT acknowledges the support of NSF grant
AST-0507325. We sincerely thank the anonymous referee whose remarks
significantly improved the presentation.\\
\end{acknowledgements}

\end{document}